# Current-phase relation of graphene Josephson junctions

C. Chialvo, I. C. Moraru, D. J. Van Harlingen, N. Mason

Department of Physics and Materials Research Laboratory, University of Illinois at Urbana-Champaign, Urbana IL 61801, USA

The current-phase relation (CPR) of a Josephson junction reveals valuable information about the microscopic processes and symmetries that influence the supercurrent. In this Letter, we present direct measurements of the CPR for Josephson junctions with a graphene barrier, obtained by a phase-sensitive SQUID interferometry technique. We find that the CPR is skewed with respect to the commonly observed sinusoidal behavior. The amount of skewness varies linearly with critical current ($I_c$) regardless of whether $I_c$ is tuned by the carrier density or temperature.

The interplay of superconductivity and the unique electronic structure of graphene leads to unusual coherence effects such as gate-tunable supercurrents [1] and specular Andreev reflection [2]. Much recent work has focused on graphene-based Josephson junctions in which theoretical [3,4] and experimental [5,6,7] studies have examined the effects of parameters such as the junction geometry and barrier thickness on the critical current. However, key information about the microscopic processes that contribute to the supercurrent can be obtained by measuring not just the magnitude of the supercurrent but also its dependence on the phase difference across the junction, characterized by the Josephson current-phase relation (CPR). The simplest models of Josephson tunneling predict a sinusoidal variation of the current with phase; deviations such as skewness have been predicted [8], but are rarely observed. It is possible to extract some information about the CPR of a junction by measuring critical current diffraction patterns, Shapiro steps, or switching current in a SQUID configuration [9], but the most definitive approach is to measure the CPR directly using a phase-sensitive interferometer technique.

In this Letter, we present experimental measurements of the CPR in Josephson junctions with a single-layer graphene barrier. The junction is incorporated into a superconducting loop coupled to a dc SQUID, which allows the junction phase to be extracted directly. We observe significant deviations from sinusoidal behavior, in particular a skewness that varies linearly as a function of carrier density and temperature. Increased skewness is observed regardless of whether the critical current is enhanced by increasing the carrier density or by lowering the temperature. We compare our results to theoretical predictions of skewed CPR in ballistic graphene, and find that the low density of states and gate tunability of graphene is likely key to the observation and characterization of the skewed CPR.

Measurements were performed on three different lateral Josephson junctions fabricated on single-layer graphene flakes, all of which exhibited a similar non-sinusoidal CPR. A typical junction is shown in Fig. 1(a). The data presented here comes from two devices, sample A ($L$=340nm, $W$=27μm) and sample B ($L$=280nm, $W$=13μm). All samples were prepared by mechanical exfoliation [10] of graphite flakes onto highly p-doped Si substrates covered by 300nm of thermally grown $SiO_2$, where the doped substrate acts as a global backgate. Samples were cleaned at a temperature of 400C in $H_2Ar$, characterized by optical imaging and atomic force microscopy, and then contacted by electron-beam evaporation of Ti(4nm)/Al(60nm)/Au(7nm). Measurements were performed in a dilution refrigerator with a base temperature of 10mK. Material properties were estimated from the normal state conductivity of identically prepared samples configured for four-point measurements [see Ref. 7 and Supplementary Material], which gave a mean free path of $l \sim 25$ nm, diffusion length $D \sim 0.0124$ m$^2$/s, and Thouless energy $\varepsilon_c = \hbar D/L^2 \sim 2.6\times10^{-4}$eV. Given $\Delta_{Al} \sim 1.8\times10^{-4}$eV ($T_c \sim 1$K), $\Delta_{Al} < \varepsilon_c$ implies that our samples are marginally in the "short-junction" transport regime, and $l < L$ implies that they are in the diffusive regime.

The circuit used to extract the CPR is shown in Fig. 1(b). The Josephson junction is connected in parallel with a fabricated planar thin film superconducting loop of inductance $L$. A current $I$ injected into the circuit divides so that the phases across the junction and the loop are equal, i.e., $\phi = 2\pi(\Phi/\Phi_0)$, where $\Phi$ is the flux in the loop and $\Phi_0$ is a flux quantum. The component of current in the loop inductor is measured by coupling the flux in the loop to a commercial dc SQUID via a filamentary superconducting flux transformer. From the circuit in Fig. 1(b), it can be seen that the current $I_J$ in the junction is given by

$$I_J(\phi) = I - I_L = I - \frac{\Phi}{L} = I - \frac{V_{SQUID}}{V_\Phi L} \qquad (1)$$

where $I_L = \Phi/L$ is the current through the inductor, and $V_{SQUID} = V_\Phi \Phi$ is the measured SQUID voltage, with $V_\Phi$ the flux transfer function. It is important to reduce the geometric inductance in series with the junction so that no significant phase winding occurs other than that across the junction; such stray inductance can distort the CPR. To achieve this, we connect the current leads and pickup loop to opposite ends of the sample as shown in Fig. 1(b). In addition, in order to maintain the circuit in the non-hysteretic regime, the condition $\beta_L = 2\pi L I_c/\Phi_0$ must be met. For our samples, $L = 3$-4nH, requiring $I_c < 100$nA to keep $\beta_L < 1$. To satisfy this constraint, $I_c$ was reduced by measuring the CPR at temperatures in the range $T = 850$mK to $890$mK. Reducing $I_c$ also ensures that the junction is in the short Josephson junction limit so that the current-phase relation is well-defined.

Figure 2(a) shows typical measurements of the SQUID voltage vs. applied current $I$, from which the CPR curves such as those presented in Fig. 3 can be extracted. The variation of $I_c$ with gate voltage $V_G$, determined from the amplitude of the CPR curves, is plotted in Fig. 2(b). Away from the Dirac point, $I_c$ increases with the magnitude of the gate voltage as expected. However, near the Dirac point, $I_c$ remains finite and exhibits anomalous structure, likely due to charges trapped nearby in the substrate [11,12]. Although the gate voltage behavior is sample dependent, the extracted CPR curves all had a similar shape. The CPR results shown were extracted for $|V_G| > 10$V, where contributions from trapped charges are not apparent.

Figure 3(a) shows the CPR vs. $V_G$ for sample A, extracted from Fig. 2(a), for $-40V < V_G < -15V$ in steps of 5V. Similar CPR curves were observed for positive gate voltages. The curves are clearly non-sinusoidal and skewed to one side, with the amount of skewness increasing with increasing magnitude of gate voltage. Figure 3(b) shows similar CPR data taken at fixed gate voltage ($V_G = -40$V) as a function of temperature. Here, the amount of skewness increases with decreasing temperature. The similar behavior of the CPR as a function of gate voltage and temperature implies that the skewness is governed by the overall value of the critical current, and not simply the distance to the Dirac point.

It is convenient at this point to parameterize the skewness by a variable $S = (2\varphi_{max}/\pi) - 1$, where $\varphi_{max}$ is the position of the maxima of the CPR; $S$ ranges from 0 to 1 as the CPR evolves from a sine wave toward a ramp wave. In Fig. 4 we plot $S$ vs. $I_c$

extracted from CPR data of sample B (similar results were obtained from sample A). The data show that $S$ increases approximately linearly with increasing $I_c$, regardless of whether $I_c$ was tuned by temperature or by gate voltage. We note two significant features of this curve. First, the linear variation of the skewness does not extrapolate to zero in the limit of zero critical current. Second, we are able to extract the CPR and measure its skewness even for $\beta_L > 1$, for which the CPR is expected to be hysteretic (see for example the red data curve in Fig. 5 for which $\beta_L = 1.35$). We attribute this to external noise which rounds the measured $V_{SQUID}$ vs. $I$ curves sufficiently to suppress the hysteresis for $\beta_L \gtrsim 1$; however, simulations of the noise rounding indicate that it does not significantly distort the shape of the CPR nor change its skewness.

The physical mechanism behind the skewness and its linear dependence on $I_c$ in our junctions is not fully understood. The CPR of a ballistic superconductor-graphene-superconductor junction, at zero temperature and in the short-junction regime, was predicted theoretically [13] using the Dirac-Bogoliubov-de Gennes (DBdG) formalism. The critical current $I_c$ is carried by Andreev bound states in the junction according to

$$I_c(\phi) = \frac{e\Delta}{\hbar} \sum_{n=0}^{\infty} \frac{T_n \sin(\phi)}{\sqrt{1 - T_n \sin^2(\phi/2)}} \quad (2)$$

where $\Delta$ is the superconducting energy gap and the $T_n$ are the transmission coefficients, which are functions of the Andreev bound state wavevectors. It can be seen from Eq. (2) that large values of $T_n$ lead to a non-sinusoidal $I_c(\phi)$. Near the Dirac point there are few Andreev bound states, so the additional contributions of exponentially-decaying evanescent states must be considered, which leads to a skewed CPR given by

$$I_c(\phi) = \frac{e\Delta}{\hbar}\left(\frac{2W}{L}\right)\left(\cos\frac{\phi}{2}\right)\operatorname{atanh}\left(\sin\frac{\phi}{2}\right). \tag{3}$$

This has a skewness of 0.255 at the Dirac point. In this model, when finite gate voltages are applied to the graphene, the number of bound states increases and dominates the supercurrent. This causes the skewness in the CPR to first increase, then oscillate due to interference effects between the bound states, and then saturate at $S \sim 0.42$, with increasing carrier density. Extensions of the DBdG to include self-consistency which allows the calculations to be carried out in the long-junction regime predict a similarly skewed CPR [14]. We find that our lowest temperature data taken near the Dirac point fit well to Eq. (3), as can be seen in the green fit curve of Fig. 5. For this fit, only the amplitude of the function is taken as a free parameter. By comparison, data taken at higher temperatures and gate voltages deviate substantially from this functional form (see the blue and red fit curves of Fig. 5, respectively). We observe skewness that increases with gate voltage as predicted, though we do not observe oscillations or a saturation of $S$. We also find that the temperature dependence of our samples is consistent with recent calculations that predict a decrease in skewness with increasing temperature for ballistic graphene junctions [15].

Although some of our data is consistent with calculations for ballistic graphene, it is not clear to what extent these calculations should apply to our diffusive junctions. A clue may lie in the comparison to predictions of skewed CPRs in superconducting constrictions and point contact junctions [16]. Both diffusive and ballistic superconducting constrictions are predicted to exhibit skewed CPRs [17]; in fact, the form of Eq. (3) is identical to that of the CPR obtained theoretically for a diffusive constriction [18]. The common factor between graphene and a constriction is a low density of conduction channels, and we believe this to be the key to observing a skewed CPR. Thus, as we observe, even a diffusive graphene junction should exhibit skewness similar to that predicted for ballistic junctions. To support this, theoretical work has shown that the temperature dependence of $I_c$ is qualitatively the same for ballistic and diffusive graphene junctions [15]. We note that skewed CPRs are rarely observed. Although some evidence for skewness has been seen in the temperature dependence of ballistic quantum point contacts [19] and ballistic InAs junctions [20], many of the predictions of skewed CPRs in constrictions, point contacts, and graphene have not previously been experimentally verified. In addition, the gate-voltage dependence of skewness has not previously been measured.

In conclusion, we have used a phase-sensitive interferometry technique to measure directly the current-phase relation of Josephson junctions with a graphene barrier. The CPR curves are non-sinusoidal, and display a skewness that increases linearly with critical current.

We thank Dan Bahr and Martin Stehno for technical assistance, and Annica Black-Schaffer, John Clem, Jozsef Cserti, and Jack Sadleir for useful discussions. This work was supported by the National Science Foundation under grant DMR09-06521, and by the Department of Energy, Division of Materials Science, under grant DE-FG02-07ER46453 through the Frederick Seitz Materials Research Laboratory at the University of Illinois.

Figure Captions:

Figure 1. (a) SEM micrograph showing two typical SQUID junctions. Sample A is labeled. Ti/Al/Au leads are in light gray, and outline of original graphene piece is shown with dotted white line. (b) Circuit used to measure the CPR, with graphene depicted in yellow and superconducting leads in blue.

Figure 2. (a) Raw data curves showing SQUID voltage vs. applied current bias, for various gate voltages, taken at T=850mK. (b) Critical current vs. gate voltage extracted from the CPR measurements at T=850mK.

Figure 3. (a) CPR for different gate voltages at T=850mK, for sample A, showing an increase in skewness with increasing carrier density. (b) CPR for different temperatures at $V_G$ = -40V, for sample A, exhibiting increasing skewness with decreasing temperature.

Figure 4. Skewness vs. $I_c$ extracted from CPR data for sample B. Solid line is least squares line fit. For each temperature shown (colored symbols), $I_c$ is shifted with gate voltage. Horizontal line marks maximum predicted skewness at the Dirac point, as discussed in the text; vertical line marks where hysteresis parameter $\beta_L$ = 1.

Figure 5. Fits of Eq. (3) to the measured CPR curves obtained at the Dirac point at T=850mK (green/middle curve) and T=880mK (blue/lower curve), and at $V_{gate}$=35V for T=850mK (red/upper curve).

**Figure 1**

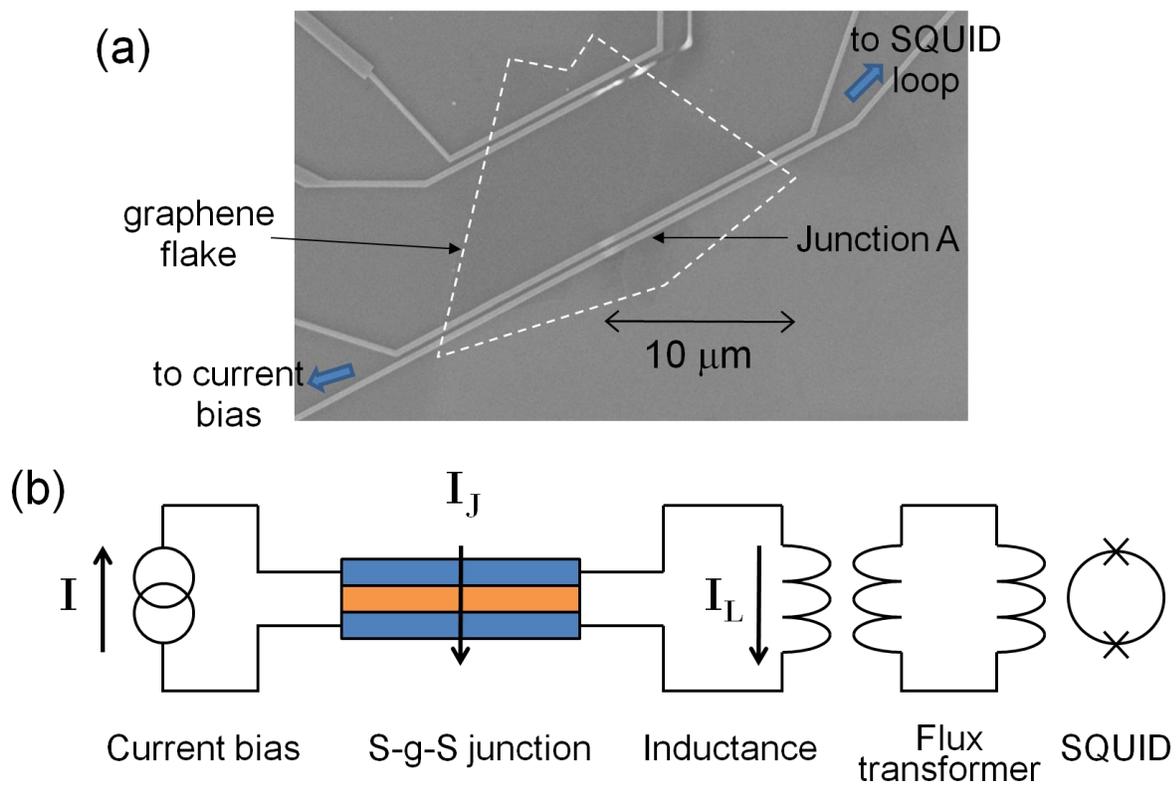

**Figure 2**

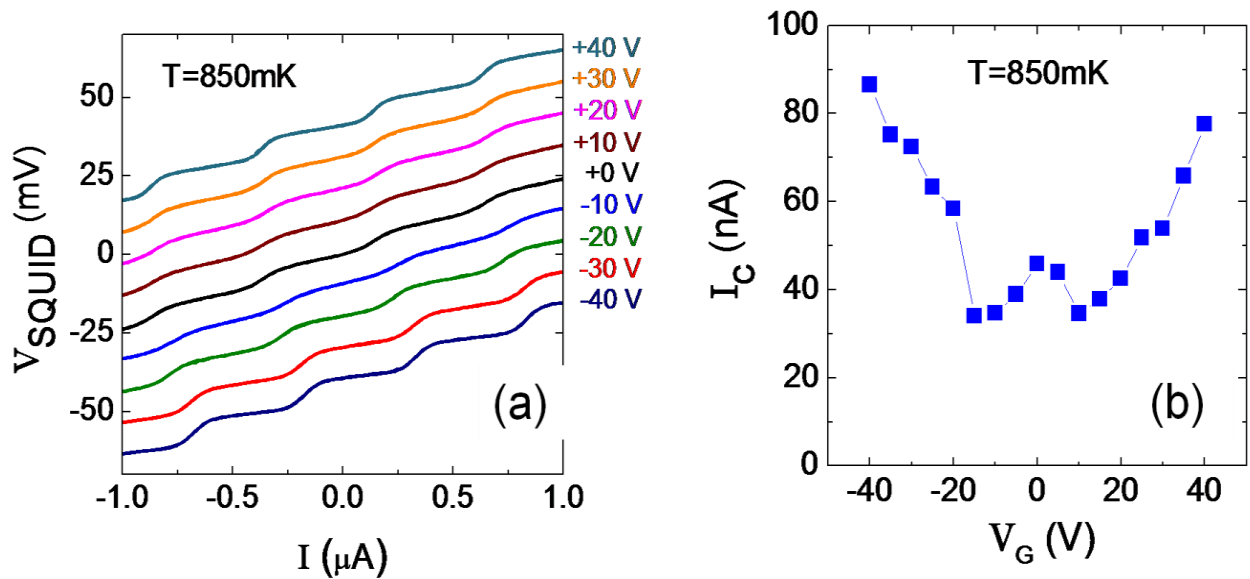

**Figure 3**

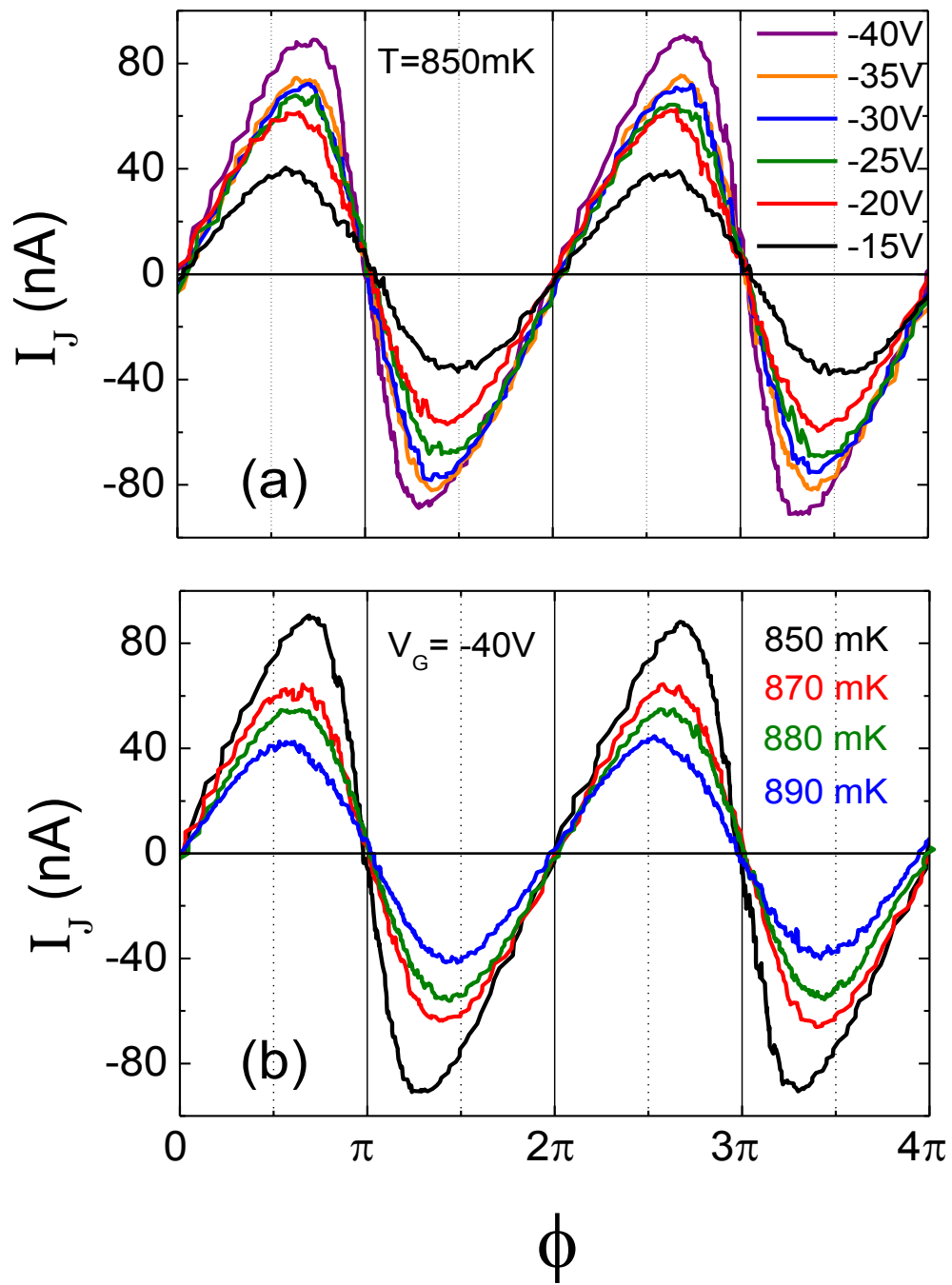

**Figure 4**

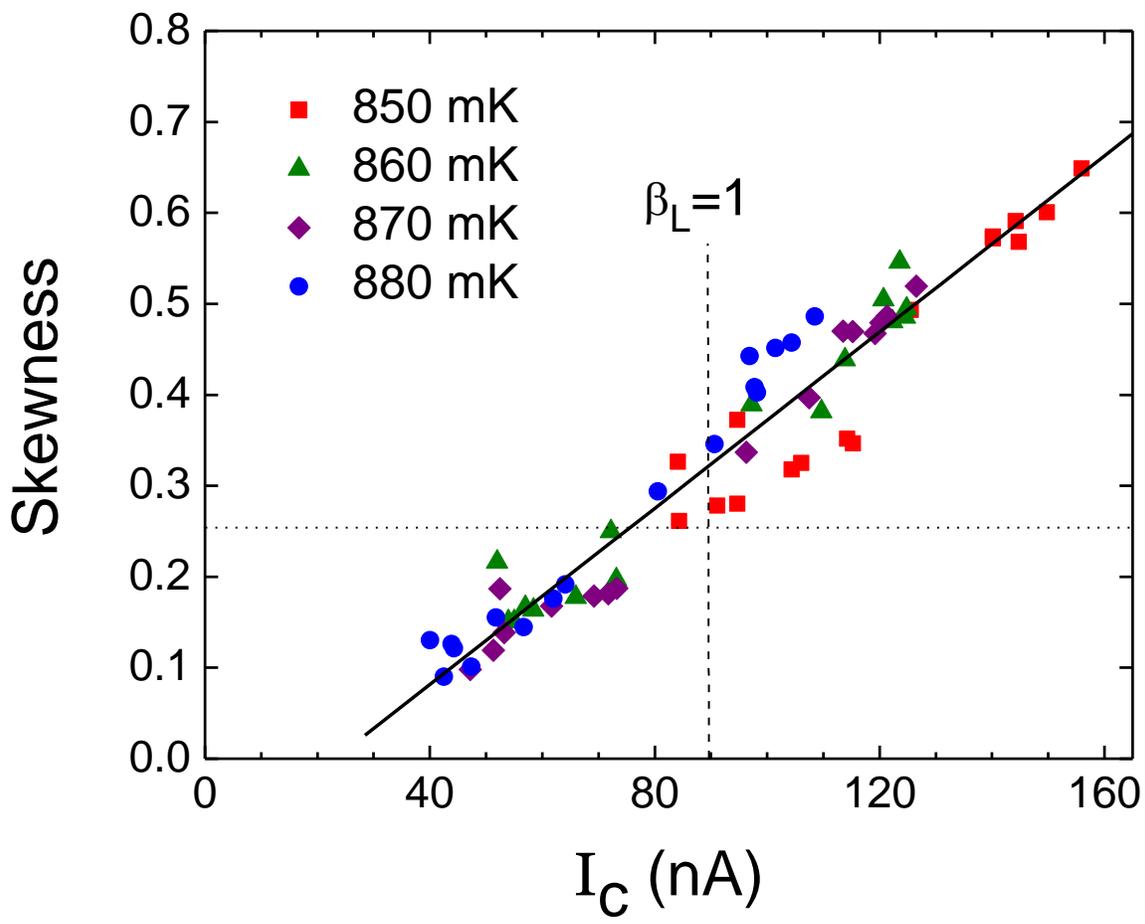

**Figure 5**

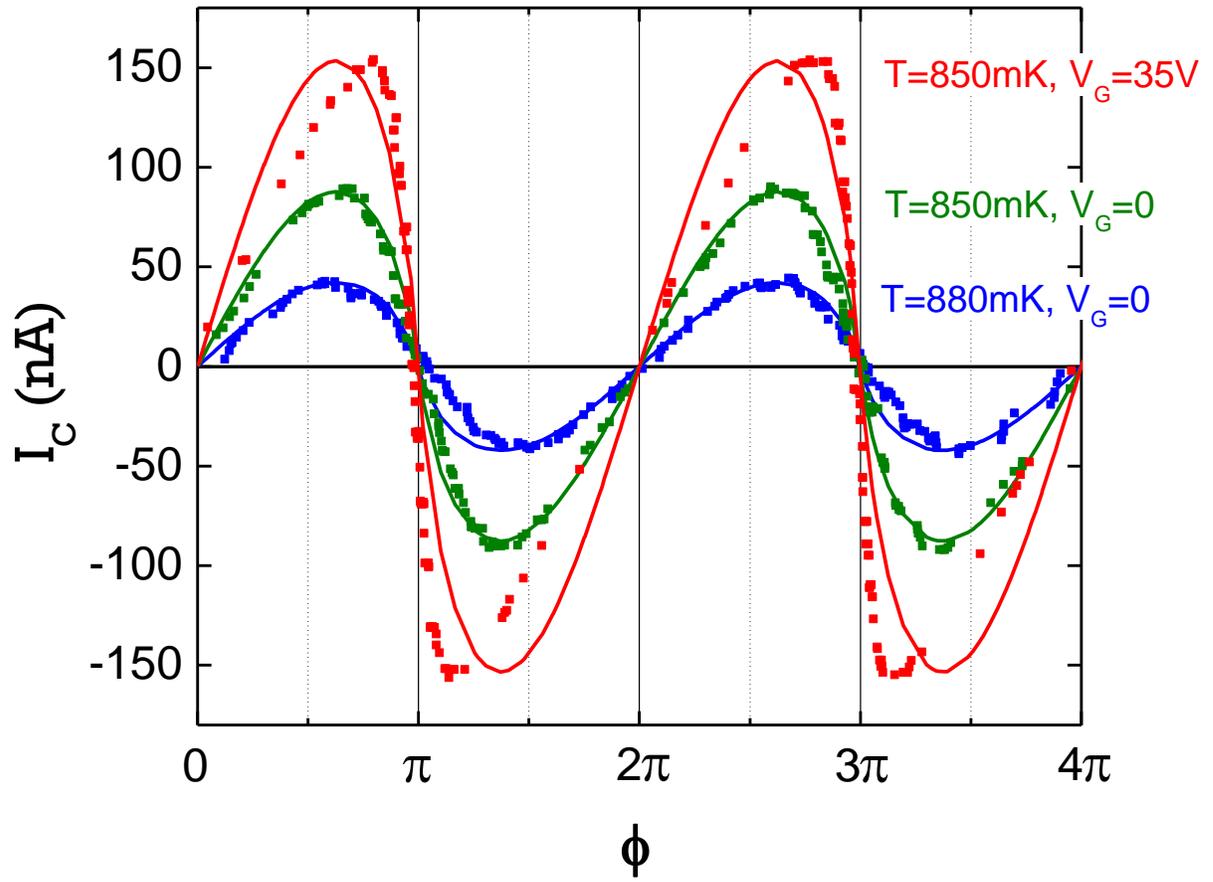